\documentclass[11pt]{article}

\usepackage{amsmath,amssymb}
\usepackage[mathscr]{eucal}
\usepackage{pgfplots}

\usepackage[normalem]{ulem}			

\DeclareMathOperator{\Tr}{Tr}
\DeclareMathOperator{\Res}{Res}
\DeclareMathOperator{\Det}{Det}
\DeclareMathOperator{\ctg}{ctg}

\newcommand{\pl}{\partial}
\newcommand{\ol}{\overline}

\newcommand{\ee}{\mathbf{e}}

\newcommand{\BB}{\mathcal{B}}
\newcommand{\DD}{\mathcal{D}}
\newcommand{\HH}{\mathscr{H}}
\newcommand{\MM}{\mathcal{M}}
\newcommand{\WW}{\mathscr{W}}
\newcommand{\RE}{\mathrm{Re}}
\newcommand{\RR}{\mathbb{R}}
\newcommand{\CC}{\mathbb{C}}

\newcommand{\OO}{\mathcal{O}}
\newcommand{\ve}{\varepsilon}
\newcommand{\ww}{w}	
\newcommand{\EE}{J}	
\newcommand{\PP}{P}	
\newcommand{\bb}{S} 

\renewcommand{\vec}[1]{\boldsymbol{#1}}
\newcommand{\braket}[2]{\langle #1 | #2 \rangle}
\newcommand{\mel}[3]{\langle #1 | #2 | #3 \rangle}

\begin{document}

\centerline{\textbf{
Singular perturbations of a free quantum field Hamiltonian
}}

\bigskip

\centerline{\textbf{T.~A.~Bolokhov}}

\bigskip

\centerline{\it St.\,Petersburg Department of V.\,A.\,Steklov Mathematical Institute}
\centerline{\it Russian Academy of Sciences}
\centerline{\it 27 Fontanka, St.\,Petersburg, Russia 191023}


\bigskip
\centerline{Abstract}
    We study solutions of the functional eigenstate equation
    of a free quantum field Hamiltonian.
    Admissible solutions are to have a finite norm and finite eigenvalues
\textit{w.\,r.\,t.} the norm and eigenvalue of the ground state of the free
    theory.
    We show that in the simple cases of a scalar field and of a vector field in the Coulomb gauge
    the admissible eigenstates exist and possess negative energy.
    The functionals can be treated as infinite-dimensional counterparts
    of the eigenfunctions of the theory of singular perturbations of
    differential operators, and can be deployed for construction
    of the renormalized states of models with asymptotic freedom.

\bigskip


\section*{Introduction}
    The Hamiltonian of the free quantum field is an infinite-dimensional
    analogue of the Hamiltonian of the harmonic oscillator.
    Its action is the sum of two terms: kinetic -- the sum of variational derivatives, and
    potential -- the quadratic form of the Laplacian
\begin{equation}
\label{HB}
    \HH = - \int_{\RR^{3}} d^{3}x
	\frac{\delta}{\delta B_{\vec{x}}^{k}}
	\frac{\delta}{\delta B_{\vec{x}}^{k}} 
    + \int_{\RR^{3}} d^{3}x \,\bigl(
	\frac{\pl B_{\vec{x}}^{k}}{\pl x_{l}}\bigr)^{2} .
\end{equation}
    In order to define an operator of the quantum theory one also has
    to fix the boundary conditions, or to provide the set of its eigenstates.
    For the Hamiltonian of the free field such a set is generated
    by the the ground state (vacuum), which is the Gaussian functional
    of the quadratic form of the square root of the Laplacian
\begin{equation}
\label{OmD}
    \Omega_{\Delta}(B) = \exp\{-\frac{1}{2}Q_{\Delta}(B)\} ,
\end{equation}
    where
\begin{equation*}
    Q_{\Delta}(B) = (B,\Delta^{1/2}B),\quad 
    \Delta = - \frac{\pl^{2}}{\pl x_{l}^{2}} .
\end{equation*}
    Besides the free quantum theory, expression
(\ref{HB})
    emerges as a result of renormalization (running to zero of the coupling
parameter)
    of certain interacting and (or) self-interacting Hamiltonians.
    Under specific conditions those systems exhibit the phenomenon of asymptotic freedom.
    One of the consequences of the latter is the asymptotic
    approach of the eigenstates to those of the free theory,
    with the increase of energy.
    We propose that if we are able to construct the set of eigenstates
    of operation
(\ref{HB})
    distinct from the set generated by
(\ref{OmD}),
    then we encounter a model of some asymptotically free theory.

    The finite-dimensional counterparts of the suggested
    construction are the
    systems in the theory of singular perturbations of differential
    operators
\cite{BF},
\cite{AK}.
    On one side this theory describes a set of eigenstates
    of the free-particle Hamiltonian with non-trivial boundary conditions.
    On the other side, this set can be obtained as a result of running
    to zero of the coupling parameter of an interacting Hamiltonian,
    for example
\begin{equation*}
    H_{\delta} = - \frac{\pl^{2}}{\pl x_{l}^{2}}
	- \ve \delta(\vec{x}-\vec{x}^{0}), \quad \ve\to 0
\end{equation*}
    or
\begin{equation*}
    H_{x^{-2}} = - \frac{\pl^{2}}{\pl x_{l}^{2}}
	- \frac{\ve}{|\vec{x}-\vec{x}^{0}|^{2}} ,\quad \ve\to 0.
\end{equation*}
    One may note that in the latter example operator
$ H_{x^{-2}} $
    can be defined for a finite
$ \ve $
    of any sign (for the 3-dimensional space),
    while in the former case operator
$ H_{\delta} $
    requires regularization. 
    Nevertheless, both of them lead to the same set of self-adjoint
    extensions of some symmetric operator having the action
    of the Laplacian. In this work we are trying to guess
    the result of a similar effect taking place in the case of the free
    field Hamiltonian.
    Namely, we look for solutions of the
    eigenstate equation of the free Hamiltonian,
\begin{equation}
\label{HEigen}
    \HH \Omega_{\MM} = E \Omega_{\MM}
\end{equation}
    in terms of Gaussian functionals with a suitable kernel
$ Q_{\MM}(\vec{x},\vec{y}) $
\begin{equation}
\label{OmM}
    \Omega_{\MM}(B) = \exp\{-\frac{1}{2}Q_{\MM}(B)\}
	= \exp\{-\frac{1}{2}\int_{\RR^{3}} B_{\vec{x}}
	Q_{\MM}(\vec{x},\vec{y}) B_{\vec{y}}\,d^{3}x\, d^{3}y\} .
\end{equation}
    It is natural to require that this equation holds at least
    for functionals
    defined on the
    set of one-time differentiable functions on space
$ \RR^{3} $.
    And to require that the following conditions
    be satisfied:
\begin{enumerate}
\item Finiteness of the norm of the state
$ \Omega_{\MM} $
    expressed,
    using the methods of the functional integration,
    in terms of the
    norm of the ground state
$ \Omega_{\Delta} $
    of the free field.
\label{PageCond}
\item Finiteness and realvaluedness of the difference of eigenvalues of
$ \Omega_{\MM} $
    and of the ground state.
    In order to ensure the unitarity of the model one is also
    to require realvaluedness of energies of all possible excitations of 
(\ref{OmM}).
\end{enumerate}
    In what follows,
    we refer to functionals (states) satisfying the above
    conditions as \emph{admissible} functionals.

    The construction of solutions of equation
(\ref{HEigen})
    in the terms of the square roots of the extensions of the quadratic
    forms has been discussed in
\cite{TBA}.
    In the present work we, however, propose that the quadratic form
$ Q_{\MM} $
    does not just depend on a single point, but rather
    that it depends on several selected points
$ \vec{x}_{1}, \ldots \vec{x}_{N} $
    of the 3-dimensional space. It can also depend on a matrix
$ \MM $,
    which relates the coefficients of divergences of functions
$ B_{\vec{x}} $
    from the domain of
$ Q_{\MM} $.
    Points
$ \vec{x}_{1},\ldots\vec{x}_{N} $
    can be naturally related to creation operators of the
    field
$ \psi $ that was interacting with
$ B $
    before the renormalization.
    It is also natural to treat matrix
$ \MM $
    as a function of coordinates and quantum numbers of
$ \psi $
    and to assume that its particular form is defined in the process of
    renormalization.

    The presented technique can be interpreted in terms of the second
    quantization of a 3-dimensional field with a multi-point singular
    perturbation and boundary conditions which relate the coefficients of divergences
    at different points with each other.
    From that point of view, the above conditions of admissibility select
    models which correspond to Hamiltonian
(\ref{HB})
    with eigenstates that can be described as excitations
    of normalizable ground state.

    The main calculations in this work are done for the case of
    a transverse (solenoidal) vector field (a vector field in the Coulomb gauge)
\begin{equation*}
    B_{\vec{x}}^{k} :\quad \frac{\pl}{\pl x_{k}} B_{\vec{x}}^{k} = 0,
\end{equation*}
    while for a somewhat simpler case of a scalar field the answers are
    given without explanation.
    We show that at least for
$ N=2 $
    there exist non-trivial admissible solutions of equation
(\ref{HEigen})
    with negative energy \textit{w.\,r.\,t.} the energy of the ground state
    of the free field.
    This means that from energy considerations, these states appear to be
    more favoured than the ground state of the free theory.
    Despite the fact that energy is a dimensionful quantity,
    the solutions in question initially possess a scale, namely the typical
    distance between points
$ \vec{x}_{1},\ldots\vec{x}_{N} $.
    As a consequence, the set of admissible solutions is described in terms of
    the real part of a certain projective space,
    \textit{i.\,e.} as a dimensionless quantity,
    while the scale of energies of these solutions is generated by means of the plain factor of
    inverse power of the distance.
    We show that at a fixed distance
    this set, parametrized by
$ \MM $,
    possesses connected components with continuously
    varying energy.
    As long as matrix
$ \MM $
    can itself depend on the distance between the points,
    this opens up a way to exploit this model for a description
    of systems with dimensional transmutation.
    
\subsection*{Terms and notations}
    We are using the following terms and notations.
    Repeating indices are summed upon. The scalar product in the
    round brackets denotes either an integration over the 3-dimensional
    space and summation over the indices, or just a plain summation 
\begin{equation*}
    (A,B) = \sum_{k}\int_{\RR^{3}} \bar{A}_{\vec{x}}^{k} B_{\vec{x}}^{k}
	\, d^{3}x ,
    \quad (\xi_{1},\xi_{2}) = \sum_{m} \bar{\xi}_{1}^{m} \xi_{2}^{m}
\end{equation*}
    By Laplace operator
$ \Delta $
    we imply a self-adjoint operator with the action
\begin{equation*}
    \Delta = - \frac{\pl^{2}}{\pl x_{l}^{2}} ,
\end{equation*}
    which is defined on the closure of the set of smooth (scalar or transverse) functions on
$ \RR^{3} $.
    We refer to pure action
(\ref{HB})
    as the free Hamiltonian,
    while the Hamiltonian of the free field is operator
(\ref{HB})
    defined on a set generated by the ground state
(\ref{OmD}).
    We direct the branch cut of the square root along
    the positive semi-axis
\begin{equation*}
    \ol{\sqrt{\lambda}} = - \sqrt{\bar{\lambda}}, \quad
	    0 < \arg \lambda < 2\pi ,
\end{equation*}
    and choose the positive branch
\begin{equation*}
    \sqrt{\rho} = \sqrt{\rho+i0} > 0 ,\quad \rho >0.
\end{equation*}
    The
$ \pm i0 $
    notation stands for taking the limit
$ \varepsilon \to \pm 0 $
    in an analogous expression containing
$ \pm i\varepsilon $ instead,
\begin{equation*}
    \sqrt{\rho\pm i0} = \lim_{\varepsilon\to \pm 0} \sqrt{\rho\pm i\varepsilon}.
\end{equation*}

\subsection*{Structure of the work}
    This exposition consists of the following parts.
    Sections 1--3 represent some well-known results from the quantum field
    theory and the theory of linear operators in Hilbert spaces.
    In Section 1 we derive an equation for the kernel of the quadratic
    form of the Gaussian functional and give its formal solutions in terms
    of extensions of closable semi-bounded quadratic forms.
    In Section 2 we provide expressions for the resolvents
    of self-adjoint extensions of symmetric operator, parametrized by
    matrices of boundary conditions.
    In Section 3
    the admissibility conditions are rewritten in terms of the integrals of
    the resolvents and their derivatives by use of the functional calculus.
    In Section 4 the convergence of these integrals is reformulated by means of
    restrictions on matrices of boundary conditions.
    In the last section we calculate the eigenvalues of the admissible eigenstates
    corresponding to matrices of a special type.

\section{Solutions of the eigenstate equations}
\subsection{Equations for the kernel of the Gaussian functional}
    Let us denote
$ q(B) $
    to be the quadratic form of the Laplace operator
\begin{equation*}
    q(B) = \int_{\RR^{3}} d^{3}x \,\bigl(
	\frac{\pl B_{\vec{x}}^{k}}{\pl x_{l}}\bigr)^{2} 
\end{equation*}
    and write down the action of Hamiltonian 
$ \HH $
    on the Gaussian functional
(\ref{OmM})
\begin{multline}
    \HH \exp\{-\frac{1}{2}Q(B)\} = \\
    = - \int_{\RR^{3}} d^{3}x
	\frac{\delta}{\delta B_{\vec{x}}^{k}}
	\frac{\delta}{\delta B_{\vec{x}}^{k}}
	\exp\{-\frac{1}{2}Q(B)\} 
	    + q(B) \exp\{-\frac{1}{2}Q(B)\} =\\
\label{HO}
    = \Bigl(- \int_{\RR^{3}} d^{3}x
	\frac{\delta Q(B)}{\delta B_{\vec{x}}^{k}}
	\frac{\delta Q(B)}{\delta B_{\vec{x}}^{k}}
    + \int_{\RR^{3}} d^{3}x 
\frac{\delta^{2} Q(B)}{\delta B_{\vec{x}}^{k} \delta B_{\vec{x}}^{k}}
    + q(B) \Bigr)
	\exp\{-\frac{1}{2}Q(B)\} .
\end{multline}
    Expressing quadratic form
$ Q(B) $
    in terms of the kernel
$ Q^{kl}(\vec{x},\vec{y}) $
\begin{equation*}
    Q(B) = \int_{\RR^{3}} d^{3}x\, d^{3}y\, B_{\vec{x}}^{k}
	Q^{kl}(\vec{x},\vec{y}) B_{\vec{y}}^{l} ,
\end{equation*}
    equation
(\ref{HO})
    is rewritten as follows
\begin{multline*}
    \HH \exp\{-\frac{1}{2}Q(B)\} = \Bigl(
    -\frac{1}{2}\int_{\RR^{3}} B_{\vec{y}}^{l}\bigl(
	Q^{kl}(\vec{x},\vec{y}) Q^{kl'}(\vec{x},\vec{z}) + \\
	+ Q^{lk}(\vec{y},\vec{x}) Q^{kl'}(\vec{x},\vec{z})
    \bigr) B_{\vec{z}}^{l'} \,d^{3}x\,d^{3}y\,d^{3}z
	+ \Tr Q + q(B) \Bigr)	\exp\{-\frac{1}{2}Q(B)\} ,
\end{multline*}
    where we have denoted
\begin{equation*}
    \Tr Q = \int_{\RR^{3}} d^{3}x\, Q^{kk}(\vec{x},\vec{x}) .
\end{equation*}
    In order for functional
$ \exp\{-\frac{1}{2}Q(B)\} $
    to be an eigenstate of operator
$ \HH $
    with eigenvalue
$ \Tr Q $,
    the following equation has to hold
\begin{multline}
\label{QBrel}
    \frac{1}{2}\int_{\RR^{3}} B_{\vec{y}}^{l}\bigl(
	Q^{kl}(\vec{x},\vec{y}) Q^{kl'}(\vec{x},\vec{z}) 
	+ Q^{lk}(\vec{y},\vec{x}) Q^{kl'}(\vec{x},\vec{z})
    \bigr) B_{\vec{z}}^{l'} \,d^{3}x\,d^{3}y\,d^{3}z =\\
    = \int_{\RR^{3}} d^{3}x \,\bigl(
	\frac{\pl B_{\vec{x}}^{k}}{\pl x_{l}}\bigr)^{2} .
\end{multline}
    We are going to seek functional
$ \exp\{-\frac{1}{2}Q(B)\} $
    as an expression with a symmetric kernel
$ Q^{kl}(\vec{x},\vec{y}) = Q^{lk}(\vec{y},\vec{x}) $,
    so that equation
(\ref{QBrel})
    can be simplified as
\begin{equation}
\label{QBr}
    \int_{\RR^{3}} B_{\vec{y}}^{l}
	Q^{lk}(\vec{y},\vec{x}) Q^{kl'}(\vec{x},\vec{z})
    B_{\vec{z}}^{l'} \,d^{3}x\,d^{3}y\,d^{3}z
    = \int_{\RR^{3}} d^{3}x \,\bigl(
	\frac{\pl B_{\vec{x}}^{k}}{\pl x_{l}}\bigr)^{2} .
\end{equation}
    Here and in what follows we assume that both function
$ B_{\vec{x}}^{k} $
    and kernel
$ Q^{kl}(\vec{x},\vec{y}) $
    are transverse in variables
$ (\vec{x},k) $, $ (\vec{y},l) $
    and may not be wrapped with the projector operators.

\subsection{The ground state of the free field}
    Equation
(\ref{QBr})
    evidently holds for 
$ Q $
    as the kernel of the positive square root of the Laplace operator
\begin{equation}
\label{QD}
    Q_{\Delta}(\vec{x},\vec{y}) = \Delta^{1/2}(\vec{x},\vec{y}) .
\end{equation}
    The ground state
\begin{equation}
\label{expD}
    \Omega_{\Delta} = \exp\{-\frac{1}{2}Q_{\Delta}(B)\} 
\end{equation}
    and its excitations are the eigenstates of the Hamiltonian
    of the free quantum field.
    The eigenvalue of
(\ref{HO})
    is proportional to the integral
\begin{equation*}
    \Tr Q_{\Delta} \sim \int_{\RR^{3}}d^{3}x
	\int_{0}^{\infty} \lambda \,d\lambda ,
\end{equation*}
    which is a divergent quantity, to be subtracted from the energy
    of the excitations.
    Indeed, a quantum state formed by 
$ \Omega_{\Delta} $
    with a polynomial coefficient
$ a(B) $
    of power
$ M $
    can be decomposed as a sum (integral)
\begin{equation}
\label{aB}
    a(B) \Omega_{\Delta} 
    = \sum_{m=0}^{M} \int dp_{1}\ldots dp_{m} \,
	a^{m}_{\sigma_{1}\ldots\sigma_{m}}(\vec{p}_{1},\ldots \vec{p}_{m})
	\BB_{\vec{p}_{1}}^{\sigma_{1}} \ldots  \BB_{\vec{p}_{m}}^{\sigma_{m}}
    \Omega_{\Delta} ,
\end{equation}
    where
$ \BB_{\vec{p}}^{\sigma} $
    are the creation operators of a boson with momentum
$ \vec{p} $ 
    and polarization
$ \sigma $.
    Construction of these creation operators involves diagonalizing
    kernel
(\ref{QD})
    via Fourier transform, so that
$ \BB_{\vec{p}}^{\sigma} $
    commutes with
$ \HH $
    in the following way
\begin{equation*}
    \HH \BB_{\vec{p}}^{\sigma} = 
    \BB_{\vec{p}}^{\sigma} (\HH + |p|).
\end{equation*}
    This yields the relation
\begin{equation*}
    \HH	\,\BB_{\vec{p}_{1}}^{\sigma_{1}} \ldots  \BB_{\vec{p}_{m}}^{\sigma_{m}}
    \Omega_{\Delta}
    = \bigl(\sum_{k=1}^{m}|p_{k}| + \Tr Q_{\Delta} \bigr)
    \Omega_{\Delta} 
\end{equation*}
    and
\begin{equation*}
    \HH \, a(B) \Omega_{\Delta} 
    = \sum_{m=0}^{M} \int dp_{1}\ldots dp_{m} \,
	a^{m}_{\sigma_{1}\ldots\sigma_{m}}(\vec{p}_{1},\ldots \vec{p}_{m})
	\bigl(\sum_{k=1}^{m}|p_{k}| + \Tr Q_{\Delta} \bigr)
    \Omega_{\Delta} ,
\end{equation*}
    which means that the action of Hamiltonian
$ \HH_{B} $
    on state
(\ref{aB})
    results in multiplying by
$ \Tr Q_{\Delta} $
    and adding a number of finite terms
(which depend on coefficients $ a^{m}_{\sigma_{1}\ldots\sigma_{m}} $).

    The methods of functional integration
\cite{ZJ}
    allow us to endow the functionals of type
(\ref{aB})
    with a scalar product.
    Let us define the latter as a functional integral over the set of fields
$ B_{\vec{x}} $,
    from the domain of
$ Q_{\Delta}(B) $
\begin{equation*}
    \braket{\Omega_{1}}{\Omega_{2}} =C \int \bar{\Omega}_{1}(B) \Omega_{2}(B) \,
    \prod_{\vec{x},k} \delta B_{\vec{x}}^{k}
\end{equation*}
    and fix the norm as
\begin{equation*}
    \braket{\Omega_{\Delta}}{\Omega_{\Delta}}
    = C \int \exp\{-Q_{\Delta}(B)\}
	\prod_{\vec{x},k} \delta B_{\vec{x}}^{k} = 1 .
\end{equation*}
    Then, assuming that functional integration is invariant
\textit{w.\,r.\,t.}
    a shift of its variable
\begin{equation*}
    \mel{\Omega_{\Delta}}{
\exp\{2\int A_{\vec{x}}^{k} B_{\vec{x}}^{k} \,d^{3}x\} }{\Omega_{\Delta}}
= \exp\{\int_{\RR^{3}} A_{\vec{x}}^{k} Q_{\Delta kl}^{-1}(\vec{x},\vec{y})
    A_{\vec{y}}^{l} \,d^{3}x\,d^{3}y\} ,
\end{equation*}
    the scalar product of states of type
(\ref{aB})
    can be calculated in the following way
\begin{multline*}
    \braket{a_{1}(B)\Omega_{\Delta}}{a_{2}(B)\Omega_{\Delta}}
    = \mel{\Omega_{\Delta}}{\bar{a}_{1}(B) a_{2}(B)}{\Omega_{\Delta}} =\\
    = \bar{a}_{1}(\frac{1}{2}\frac{\delta}{\delta A})
    a_{2}(\frac{1}{2}\frac{\delta}{\delta A})
\exp\{\int_{\RR^{3}} A_{\vec{x}}^{k}
Q_{\Delta kl}^{-1}(\vec{x},\vec{y})
    A_{\vec{y}}^{l} \,d^{3}x\,d^{3}y\} \Bigr|_{A=0}.
\end{multline*}
    Here
$ Q_{\Delta kl}^{-1}(\vec{x},\vec{y}) $
    is the kernel of the operator inverse to
$ \Delta^{1/2} $.

\subsection{Other solutions of the equation for the kernel of quadratic form}
    One can discover that equation
(\ref{QBr})
    admits other solutions than the kernel of the square root of the Laplacian.
    We have intentionally written that equation
    not as an equality of operators,
    but rather as the equality of the values of quadratic forms.
    This allows us to take advantage of the existence of various
    quadratic
    forms with equal values on a common set (intersection)
    of their domains.
    Let us denote by
$ \DD[\Delta] $
    the set of continuous one-time differentiable functions on
$ \RR^{3} $.
    The closure
$ \bar{\DD}[\Delta] $
    of this set, in the norm induced by the positive quadratic form
$ q(B) $,
    preserves the transversality condition.
    We may assume that functionals of the quantum theory are initially
    defined on the set
$ \DD[\Delta] $
    and then continuously extended to
$ \bar{\DD}[\Delta] $.
    The quadratic form
$ q(B) $
    of the Laplace operator on the transverse subspace admits
    closable semi-bounded extensions.
    That is, there exist quadratic forms
\begin{equation*}
    q_{\MM}(B): \quad \DD_{\MM} \to \CC ,
\end{equation*}
    such that
\begin{equation*}
    q_{\MM}(B) = q(B), \quad B\in \DD[\Delta] \subset \DD_{\MM} ,
\end{equation*}
    and the domains
$ \DD_{\MM} $
    are strictly greater than
$ \DD[\Delta] $.
    In what follows we will parametrize the extensions of interest
    by a set of points
$ \vec{x}_{1},\ldots \vec{x}_{N} $
    from
$ \RR^{3} $
    and by a symmetric matrix
$ \MM $
    of size
$ 3N \times 3N $.

    Theorem VIII.15 from
\cite{RS}
    states that for any semi-bounded closable quadratic form
$ q_{\MM}(B) $
    there exists a self-adjoint operator
$ T_{\MM} $
    such that
\begin{equation*}
    q_{\MM}(B) = (B,T_{\MM}B), \quad
	B\in \DD(T_{\MM}) \subset \DD_{\MM}.
\end{equation*}
    This relation holds on the domain of operator
$ \DD(T_{\MM}) $,
    and then, using the semi-boundedness of
$ q_\MM(B) $,
    is extended to its
$ q_{\MM} $-norm closure
$ \DD_{\MM} $,
    which is referred to in the literature as
$ \DD[T_{\MM}] $.
    The spectral decomposition of operator
$ T_{\MM} $
    allows us to construct its functional calculus and, in particular,
    the square root
$ T_{\MM}^{1/2} $.
    That is, to construct an operator satisfying the equation
\begin{equation*}
    (T_{\MM}^{1/2}B, T_{\MM}^{1/2}B) = q_{\MM}(B),
\end{equation*}
    which, when restricted to the set
$ \DD[\Delta] $,
    gives a solution of
(\ref{QBrel}).
    Thus, the quadratic form
$ Q_{\MM}(B) $
    of operator
$ T_{\MM}^{1/2} $
\begin{equation}
\label{QAM}
    Q_{\MM}(B) = (B,T_{\MM}^{1/2}B) , \quad B\in\DD(T_{\MM})
\end{equation}
    can be exploited in construction of eigenstates
\begin{equation}
\label{OmMT}
    \Omega_{\MM} = \exp\{-\frac{1}{2}Q_{\MM}(B)\}
	= \exp\{-\frac{1}{2}\int_{\RR^{3}} B_{\vec{x}}
	T_{\MM}^{1/2}(\vec{x},\vec{y}) B_{\vec{y}}\,d^{3}x\, d^{3}y\}
\end{equation}
    of operator
$ \HH $
\begin{equation*}
    \HH \, \Omega_{\MM} = \Tr T_{\MM}^{1/2} \,\Omega_{\MM} .
\end{equation*}
    Here one may note that, despite the fact that the values of the form
$ q_{\MM} $
    on the set
$ \DD[\Delta] \subset \DD[T_{\MM}] $
    coincide with those of 
    the Laplace operator
$ \Delta $,
    the action of the quadratic form
$ Q_{\MM}(B) $
    of the operator
$ T_{\MM}^{1/2} $
    on
$ \DD[\Delta] $
    significantly differs from the action of the form
$ Q_{\Delta}(B) $ therein.
    This means that the Gaussian functional
$ \Omega_{\MM} $
    represents a new solution of the
    eigenstate equation
(\ref{HEigen}),
    distinct from the free-field one.

\section{Resolvents of self-adjoint extensions of symmetric operator}
\subsection{Krein's resolvent formula}
    Let us focus more attention on the closable extensions of the
    quadratic form of Laplace operator
$ q(B) $.
    In doing this, we will be using the Birman-Krein-Vishik (BKV) theory
\cite{BKV},
\cite{Krein}
    describing the structure of
    self-adjoint extensions of closable semi-bounded
    symmetric operators.
    We assume that the asymptotically free interaction of field
$ B $
    with other sources, after renormalization, reduces just to certain
    non-trivial boundary conditions.
    More precisely, we propose, that the creation operators of
    the second field, 
    located at points
$ \vec{x}_{1}, \dots \vec{x}_{N} $,
    lead to a possibility of an infinite growth of field
$ B $
    at these points.
    And that the main coefficients of growth
    are related by matrix
$ \MM $
    in some way which depends on the wave function of the second field.

    Let us consider an operator
$ \Delta_{\{\vec{x}_{n}\}} $
    coinciding, by action, with the Laplace operator, but defined
    on the set    
$ \WW_{\{\vec{x}_{n}\}} $
    of smooth transverse functions vanishing at points
$ \vec{x}_{1}, \dots \vec{x}_{N} $
    with their first derivatives.
    This operator is symmetric and possesses non-trivial deficiency
    indices
$ (3N,3N) $.
    Self-adjoint extensions of operators of this kind
    are conveniently described in terms
    of boundary values, by the theory of vectors of boundary values
(the details can be found in \cite{GG}).

    The construction of resolvents in BKV theory extensively uses the
    notion of the analytic deficiency vector\footnote{Here the term \emph{vector}
is used in the mathematical sense, to denote an element of the
finite-dimensional deficiency subspace. From the physical point of view,
this object is a tensor, because, besides index
$ m $, which enters
$ \alpha $,
as a transverse function it possesses an implicit
vector index.}\!.
    Let function
$ D_{\lambda}^{\alpha}(\vec{x}) $,
$ \alpha = (n,m) $
    represent an element of the deficiency subspace of operator
$ \Delta_{\{\vec{x}_{n}\}} $
    at point
$ \bar{\lambda} \in \CC $
    not resting on the positive semi-axis. That is,
$ D_{\lambda}^{\alpha}(\vec{x}) $
    satisfies the equation
\begin{equation*}
    \bigl(D_{\lambda}^{\alpha}, (\Delta - \bar{\lambda}) h \bigr) = 0 ,
	\quad h \in \WW_{\{\vec{x}_{n}\}}.
\end{equation*}
    Let it also be analytic in argument
$ \lambda $
    and satisfy the relation
\begin{equation}
\label{Deq}
D_{\lambda} = D_{\mu} + (\lambda-\mu) R_{\lambda} D_{\mu},
\end{equation}
    where
$ R_{\lambda} $
    is the resolvent of a certain distinguished self-adjoint extension of
$ \Delta_{\{\vec{x}_{n}\}} $.
    Since we are only aware of a few of those, we take one
    to be the resolvent of the Laplace operator with the kernel
\begin{equation*}
    R_{\lambda}^{kk'}(\vec{x},\vec{y})
	= \frac{e^{i\sqrt{\lambda}|\vec{x}-\vec{y}|}}{4\pi |\vec{x}-\vec{y}|}
	    \delta_{kk'} .
\end{equation*}
    It is not hard to see that the transverse subspace is invariant
    \textit{w.\,r.\,t.} the action of the above expression,
    and thus we are going
    to handle it as though it was wrapped with transverse projectors.

    In
\cite{TB2}
    it was shown that the analytic deficiency vectors for the transverse
    symmetric operator
$ \Delta_{\{\vec{x}_{n}\}} $
    have
    the following form
\begin{equation}
\label{Dtr}
    D_{\lambda}^{mn}(\vec{x})
    = \vec{\pl}\times ((\vec{x}-\vec{x}_{n})\times\vec{\pl})
    \frac{\sqrt{2}w(\sqrt{\lambda}|\vec{x}-\vec{x}_{n}|)}{3\sqrt{\lambda}
	|\vec{x}-\vec{x}_{n}|} Y_{1m}(\frac{\vec{x}-\vec{x}_{n}}{
	|\vec{x}-\vec{x}_{n}|}),
\end{equation}
    where
\begin{equation}
\label{omt}
    \ww(t) = \frac{3}{2}\frac{d}{dt} \frac{e^{it}-1}{t}
= \frac{3}{2t^{2}}(it e^{it} -e^{it} +1) ,
\end{equation}
    and
$ Y_{1m} $
    are the 3-dimensional spherical functions.
    The analytic deficiency vectors for the 
    symmetric operator corresponding to the scalar Laplacian
    look as follows
\begin{equation}
\label{Dscal}
    D_{\lambda}^{n}  
	= \frac{e^{i\sqrt{\lambda}|\vec{x}-\vec{x}_{n}|}}{\sqrt{4\pi}
	|\vec{x}-\vec{x}_{n}|} .
\end{equation}
    In the theory of singular perturbations
\cite{AK}
    vectors
$ D_{\lambda}^{\alpha} $
    are introduced by means of
    the action of the resolvent of the unperturbed
    operator
$ \Delta $
    on the singular potential, which in our case is the
$ \delta $-function
    at one of points
$ \vec{x}^{n} $.
    For the scalar case
(\ref{Dscal}),
    this interpretation is obvious (up to a coefficient),
    while for the vector case it 
    is hidden by
    the projector
    on the transverse subspace.

    The work of Krein
\cite{Krein}
    shows that resolvent
$ \tilde{R}_{\lambda} $
    of an arbitrary self-adjoint extension of a symmetric operator
    can be expressed via its deficiency vectors
$ D_{\lambda}^{\alpha} $
    and the resolvent
$ R_{\mu} $
\begin{equation}
\label{KreinRes}
    \tilde{R}_{\lambda} = R_{\lambda}
	+ \bb_{\alpha\beta}(\lambda)
	D_{\lambda}^{\alpha} (D_{\bar{\lambda}}^{\beta}, \: \cdot \:) ,
\end{equation}
    provided that matrix function
$ \bb_{\alpha\beta}(\lambda) $
    satisfies the equation
\begin{equation}
\label{KreinEq}
    \bb_{\alpha\beta}^{-1}(\mu) - \bb_{\alpha\beta}^{-1}(\lambda)
	= (\lambda -\mu) (D_{\bar{\mu}}^{\alpha}, D_{\lambda}^{\beta}) .
\end{equation}
    Thus, taking different solutions of equation
(\ref{KreinEq}),
    we obtain resolvents
(\ref{KreinRes})
    of different self-adjoint extensions of symmetric operator
$ \Delta_{\{\vec{x}_{n}\}} $.

\subsection{Solutions of the equation for the Krein's formula}
    In order to derive the solutions of equation
(\ref{KreinEq})
    it is convenient to employ the theory of vectors of boundary values.
    Let us consider the operator
$ \Delta_{\{\vec{x}_{n}\}}^{*} $,
    adjoint to
$ \Delta_{\{\vec{x}_{n}\}} $.
    Its domain
$ \WW_{\{\vec{x}_{n}\}}^{*} $
    is comprised of such elements
$ d $
    for which the expression
\begin{equation*}
    (d, \Delta_{\{\vec{x}_{n}\}} h) = (d, \Delta h),
	\quad h \in \WW_{\{\vec{x}_{n}\}} 
\end{equation*}
    defines
    a bounded linear functional in
$ h $.
    As long as operator
$ \Delta_{\{\vec{x}_{n}\}} $
    is defined on the domain of functions vanishing at points
$ \vec{x}_{n} $
    with their first derivatives, the domain of
$ \Delta_{\{\vec{x}_{n}\}}^{*} $
    is wider than that of
$ \Delta_{\{\vec{x}_{n}\}} $
    by a set of functions with
    unlimited growth at points
$ \vec{x}_{n} $.
    Let us introduce
    the operators of boundary values
$ \Xi $ and
$ \Gamma $
    defined on 
$ \WW_{\{\vec{x}_{n}\}}^{*} $
    as linear maps
\begin{equation*}
    \Xi: \:  \WW_{\{\vec{x}_{n}\}}^{*} \to \CC^{3N} , \quad
    \Gamma: \:  \WW_{\{\vec{x}_{n}\}}^{*} \to \CC^{3N} ,
\end{equation*}
    annihilating the set
$ \WW_{\{\vec{x}_{n}\}} $
    and satisfying the ``Green's relation''
\begin{equation*}
    (\Delta_{\{\vec{x}_{n}\}}^{*} d ,g) - (d, \Delta_{\{\vec{x}_{n}\}}^{*} g)
    = (\Gamma d, \Xi g)- (\Xi d, \Gamma g) ,\quad
	d,g \in \WW_{\{\vec{x}_{n}\}}^{*} .
\end{equation*}
    Here the RHS involves the scalar products in the finite-dimensional space
$ \CC^{3N} $.
    Roughly speaking, operators
$ \Xi $ and
$ \Gamma $
    can be treated as vector-valued linear functionals mapping transverse
    functions into the vectors of their minus first (for 
$ \Xi $)
    and zeroth (for
$ \Gamma $)
    coefficients of power series at points
$ \vec{x}_{n} $.

    Clearly operators
$ \Xi $ and
$ \Gamma $
    are defined up to an unitary transformation in
$ \CC^{3N} $.
    Moreover in our case coefficients
$ [\Xi D_{\lambda}^{\alpha}]_{\beta} $
    do not depend on
$ \lambda $
    and can be chosen as
\begin{equation*}
    [\Xi D_{\lambda}^{\alpha}]_{\beta} = \delta_{\alpha\beta} .
\end{equation*}
    That is, the coefficients at the main singularities at points
$ \vec{x}_{n} $
    for 3 different components of
$ D_{\lambda}^{mn} $
    are equal to 1.
    We also assume that the action of
$ \Gamma $
    on 
$ D_{\lambda}^{\alpha} $
    is real and symmetric
\begin{equation*}
    \ol{[\Gamma D_{\lambda}^{\alpha}]}_{\beta}
    = [\Gamma D_{\bar{\lambda}}^{\alpha}]_{\beta} , \quad
    [\Gamma D_{\lambda}^{\alpha}]_{\beta} 
    = [\Gamma D_{\lambda}^{\beta}]_{\alpha}  
\end{equation*}
    and shall further denote
$ [\Gamma D_{\lambda}^{\alpha}]_{\beta} $
    as
$ \Gamma_{\lambda}^{\alpha\beta} $.
    Now one can show that the matrix function
$ \bb_{\alpha\beta}(\lambda) $,
    defined via the relation for its inverse
\begin{equation}
\label{bexpr}
    \bb_{\alpha\beta}^{-1}(\lambda) = \MM_{\alpha\beta} -
    \Gamma_{\lambda}^{\alpha\beta},
\end{equation}
    satisfies equation
(\ref{KreinEq}).
    Indeed
\begin{multline}
    \bb_{\alpha\beta}^{-1}(\lambda) - \bb_{\alpha\beta}^{-1}(\mu) =
    \Gamma_{\mu}^{\alpha\beta} - \Gamma_{\lambda}^{\alpha\beta} 
    = (\Xi D_{\bar{\lambda}}^{\alpha}, \Gamma D_{\mu}^{\beta})
	- (\Gamma D_{\bar{\lambda}}^{\alpha}, \Xi D_{\mu}^{\beta}) =\\
\label{GD}
	= (D_{\bar{\lambda}}^{\alpha}, \Delta_{\{\vec{x}_{n}\}}^{*}
		D_{\mu}^{\beta})
	-(\Delta_{\{\vec{x}_{n}\}}^{*} D_{\bar{\lambda}}^{\alpha},
		D_{\mu}^{\beta})
    = (\mu-\lambda) (D_{\bar{\lambda}}^{\alpha}, D_{\mu}^{\beta}) .
\end{multline}
    In order for function
$ R_{\lambda}^{\MM} $,
    constructed using
(\ref{KreinRes})
    and involving solution
(\ref{bexpr}),
    to be a resolvent of a self-adjoint operator it has to
    satisfy the condition 
\begin{equation*}
    R_{\lambda}^{\MM *} = R_{\bar{\lambda}}^{\MM} , \quad
    [R_{\lambda}^{\MM *}]^{\alpha\beta}(\vec{x},\vec{y}) =
	[R_{\bar{\lambda}}^{\MM}]^{\beta\alpha}(\vec{y},\vec{x}).
\end{equation*}
    The latter, together with the property
\begin{equation*}
    \bar{D}_{\lambda}^{\alpha} = D_{\bar{\lambda}}^{\alpha} ,
\end{equation*}
    yields
\begin{equation*}
    \ol{\bb_{\alpha\beta}(\lambda)} = \bb_{\beta\alpha}(\bar{\lambda}) ,
\end{equation*}
    and then one can conclude that matrix
$ \MM $
    from
(\ref{bexpr})
    has to be Hermitian
\begin{equation*}
    \MM^{T} = \bar{\MM}, \quad \bar{\MM}_{\alpha\beta} = \MM_{\beta\alpha} .
\end{equation*}
    It is evident that different Hermitian matrices
$ \MM $
    correspond to different resolvents
(\ref{KreinRes}),
    that is, to different self-adjoint extensions of operator
$ \Delta_{\{\vec{x}_{n}\}} $.
    In order to list the resolvents of all possible self-adjoint extensions
    one is also to take into account matrices that are infinite
    on various linear subspaces of
$ \CC^{3N} $.
    Or, more precisely, the Krein's formula should be supplemented with
    projectors
$ \PP_{\alpha}^{m} $
    onto all possible linear subspaces of dimension
$ M $,
$ 0 \leq M\leq 3N $
\begin{equation}
\label{KreinResp}
    R_{\lambda}^{\MM} = R_{\lambda}
	+ \bb_{mm'}(\lambda)\PP_{\alpha}^{m} D_{\lambda}^{\alpha}
    (\bar{\PP}_{\beta}^{m'} D_{\bar{\lambda}}^{\beta}, \: \cdot \:) ,
    \quad m,m' = 1,\ldots M,
\end{equation}
    where
\begin{equation}
\label{bexprp}
    \quad \bb_{mm'}^{-1}(\lambda) = \MM_{mm'} -
\bar{\PP}_{\alpha}^{m} \Gamma_{\lambda}^{\alpha\beta} \PP_{\beta}^{m'},
    \quad \bar{\PP}_{\alpha}^{m} \PP_{\alpha}^{m'} = \delta_{mm'} .
\end{equation}
    In particular, the Laplace operator corresponds to matrix
$ \MM $
    infinite on entire space
$ \CC^{3N} $,
    or to projector
$ \PP $
    onto the empty set, that generates function
$ \bb_{\alpha\beta}(\lambda) $
    identically equal to zero.

    It is also worth to note that the zeroes of the determinant in the RHS
    of
(\ref{bexpr}) and
(\ref{bexprp})
    generate the poles of resolvent
(\ref{KreinRes})
    and are responsible for the discrete spectrum of self-adjoint extensions.

\subsection{Boundary values and matrix
$ \MM $}
    It turns out that matrix
$ \MM $
    is convenient to
    describe the domain of the corresponding self-adjoint extension
    in the terms of boundary conditions.
    The statement is that the boundary values
    of vectors from the domain of self-adjoint extension
$ T_{\MM} $
    corresponding to resolvent
$ R_{\lambda}^{\MM} $
    defined by 
(\ref{bexpr})
    are related by matrix
$ \MM $
\begin{equation}
\label{DomTM}
\DD(T_{\MM}) = \{d \in \WW_{\{\vec{x}_{n}\}}^{*} : \: \MM\Xi d = \Gamma d \}.
\end{equation}
    Indeed, the BKV theory states that the domain of
$ T_{\MM} $
    decomposes into a direct sum of
$ \WW_{\{\vec{x}_{n}\}} $
    and the linear span of vectors
$ R_{\lambda}^{\MM} D_{\mu}^{\beta} $,
$ \mu <0 $
\begin{equation*}
    \DD(T_{\MM}) = \WW_{\{\vec{x}_{n}\}} \dotplus
	\{ R_{\lambda}^{\MM} D_{\mu}^{\beta} \}, \quad \mu < 0 .
\end{equation*}
    Operators
$ \Xi $ and
$ \Gamma $
    annihilate subspace
$ \WW_{\{\vec{x}_{n}\}} $,
    and therefore it is only necessary to check that operator
$ \MM\Xi -\Gamma $
    vanishes on vectors of type
$ R_{\lambda}^{\MM} D_{\mu}^{\beta} $.
    Consider the action of
$ \MM\Xi-\Gamma $
    on an element
$ R_{\lambda}^{\MM} D_{\mu}^{\beta} $
\begin{align}
\label{Mtr0}
    &\MM_{\alpha\alpha'} [\Xi R_{\lambda}^{\MM} D_{\mu}^{\beta}]_{\alpha'}
	- [\Gamma R_{\lambda}^{\MM} D_{\mu}^{\beta}]_{\alpha}
    = \MM_{\alpha\alpha'}[\Xi R_{\lambda} D_{\mu}^{\beta}]_{\alpha'} +\\
\nonumber
    &+ \MM_{\alpha\alpha'} [\Xi D_{\lambda}^{\gamma}]_{\alpha'}
\bb_{\gamma\gamma'}(\lambda)(D_{\bar{\lambda}}^{\gamma'}, D_{\mu}^{\beta})
	- [\Gamma R_{\lambda} D_{\mu}^{\beta}]_{\alpha}
    - \Gamma_{\lambda}^{\alpha\gamma}
\bb_{\gamma\gamma'}(\lambda)(D_{\bar{\lambda}}^{\gamma'}, D_{\mu}^{\beta})
    = \\
\label{Mtr}
    &= \MM_{\alpha\alpha'} [\Xi R_{\lambda} D_{\mu}^{\beta}]_{\alpha'} +
    \bigl( \MM_{\alpha\gamma} - \Gamma_{\lambda}^{\alpha\gamma}\bigr)
    \bb_{\gamma\gamma'}(\lambda)(D_{\bar{\lambda}}^{\gamma'}, D_{\mu}^{\beta})
	- [\Gamma R_{\lambda} D_{\mu}^{\beta}]_{\alpha} .
\end{align}
    As long as
$ R_{\lambda} $
    is the resolvent of the Laplace operator, its range contains
    no divergent functions and it is nulled by operator
$ \Xi $.
    Hence the first term in
(\ref{Mtr})
    has a zero coefficient
\begin{equation*}
    [\Xi R_{\lambda} D_{\mu}^{\beta}]_{\alpha'} = 0 .
\end{equation*}
    From definition
(\ref{bexpr})
    it follows that
\begin{equation*}
    \bigl( \MM_{\alpha\gamma} - \Gamma_{\lambda}^{\alpha\gamma}\bigr)
    \bb_{\gamma\gamma'}(\lambda) = \delta_{\alpha\gamma'} ,
\end{equation*}
    and in this way the second term in
(\ref{Mtr})
    equals to
$ (D_{\bar{\lambda}}^{\alpha}, D_{\mu}^{\beta}) $.
    The third term by means of relation
(\ref{Deq})
    can be rewritten as
\begin{align*}
    [\Gamma R_{\lambda} D_{\mu}^{\beta}]_{\alpha} & =
(\lambda -\mu)^{-1}[\Gamma (D_{\lambda}^{\beta} -D_{\mu}^{\beta})]_{\alpha}
    = \\ 
    &= (\lambda -\mu)^{-1}
\bigl((\Xi D_{\bar{\lambda}}^{\alpha}, \Gamma D_{\mu}^{\beta}) 
- (\Gamma D_{\bar{\lambda}}^{\alpha}, \Xi D_{\mu}^{\beta}) \bigr)
    = (D_{\bar{\lambda}}^{\alpha}, D_{\mu}^{\beta}) ,
\end{align*}
    and it follows that the two last terms in
(\ref{Mtr})
    cancel each other.

\section{Admissibility conditions in terms of matrix $ \MM $}
    The knowledge of the resolvent of operator
$ T_{\MM} $
    allows to construct the operator's functional calculus.
    A function
$ f(T_{\MM}) $
    of the self-adjoint operator
$ T_{\MM} $
    with a continuous spectrum on the positive semi-axis and a finite number
    of negative discrete eigenvalues
$ \rho_{m} $
    can be represented as the following expansion
\begin{equation}
\label{funcTa}
    f(T_{\MM}) = \frac{1}{2\pi i}\int_{0}^{\infty} (R_{\rho +i0}^{\MM}
	-R_{\rho -i0}^{\MM}) f(\rho)\,d\rho - \sum_{m} f(\rho_{m})
	\Res[R_{\lambda}^{\MM}] |_{\lambda=\rho_{m}} .
\end{equation}
    Here we suppose that the singularities of
$ f $ -- its poles and cuts --
    do not overlap the singularities of the resolvent.

    First, let us use expression
(\ref{funcTa})
    in order to calculate the norm of functional
(\ref{OmMT}).
    The functional integration
    of the Gaussian states gives the following
    relation
\begin{equation}
\label{NormOmOm}
    \frac{\braket{\Omega_{\MM}}{\Omega_{\MM}}}{
	\braket{\Omega_{\Delta}}{\Omega_{\Delta}}}
    = \frac{\Det\Delta^{1/4}}{\Det(\RE T_{\MM}^{1/2})^{1/2}}
    =\exp\{\frac{1}{2}\Tr(\ln\Delta^{1/2}-\ln\RE T_{\MM}^{1/2})\} .
\end{equation}
    Substitution of integral
(\ref{funcTa})
    into
(\ref{NormOmOm})
    shows that the negative eigenvalues of the
    discrete spectrum
$ \rho_{m} $
    enter via the contributions of type
\begin{equation*}
    \exp\{-\frac{1}{2}\ln\RE \sqrt{\rho_{m}}\} = \exp\{-\frac{1}{2}\ln 0\} ,
\end{equation*}
    that is as infinite factors in the norm.
    Thus one can conclude that self-adjoint extensions
$ T_{\MM} $
    with a discrete spectrum are not admissible \textit{w.\,r.\,t.}
    condition 1 on
p.~\pageref{PageCond}.
    For this reason we shall restrict ourselves to the
    operators
    with no discrete spectrum in what follows.

    Now let us employ expression
(\ref{funcTa})
    for construction of functional
$ \Omega_{\MM} $.
    By examining the integrals around a contour wrapping the positive semi-axis
    and taking 
$ f(\rho) $
    to be the square root with the cut along the negative semi-axis,
    one can show that the operator with the kernel
\begin{equation}
\label{TM2}
    T_{\MM}^{1/2}(\vec{x},\vec{y}) = \frac{1}{2\pi i}\int_{0}^{\infty} \bigl(
	R_{\rho +i0}^{\MM}(\vec{x},\vec{y})
	-R_{\rho -i0}^{\MM}(\vec{x},\vec{y}) \bigr)
    \sqrt{\rho}\,d\rho ,
\end{equation}
    satisfies the relation
\begin{align*}
    T_{\MM}^{1/2} T_{\MM}^{1/2}  
    = \frac{1}{2\pi i}\int_{0}^{\infty} (R_{\rho +i0}^{\MM} -R_{\rho -i0}^{\MM})
	\rho\,d\rho
      = T_{\MM} .
\end{align*}
    Equation
(\ref{QBr})
    requires symmetricity of kernel
(\ref{TM2}).
    Substitution of the resolvent from Krein's formula
(\ref{KreinRes}) into
(\ref{funcTa})
    brings the latter to the following form
\begin{align*}
    T_{\MM}^{1/2}(\vec{x},\vec{y}) &= \frac{1}{2\pi i}\int_{0}^{\infty} \bigl(
	R_{\rho +i0}(\vec{x},\vec{y})
	+ \bb_{\alpha\beta}(\rho+i0)D_{\rho+i0}^{\alpha}(\vec{x})
	    D_{\rho+i0}^{\beta}(\vec{y}) -\\
	&-R_{\rho -i0}(\vec{x},\vec{y})
	- \bb_{\alpha\beta}(\rho-i0)D_{\rho-i0}^{\alpha}(\vec{x})
	    D_{\rho-i0}^{\beta}(\vec{y}) \bigr)
    \sqrt{\rho}\,d\rho .
\end{align*}
    Here we omit certain indices in
$ T_{\MM}^{1/2}(\vec{x},\vec{y}) $,
$ R_{\lambda}(\vec{x},\vec{y}) $ and
$ D_{\lambda}^{\alpha}(\vec{x}) $.
    In order for kernel
$ T_{\MM}^{1/2}(\vec{x},\vec{y}) $
    to be symmetric, it is necessary that matrix
$ \MM_{\alpha\beta} $,
    which appears in
$ \bb_{\alpha\beta}(\lambda) $,
    not just be Hermitian but also symmetric.

    Now, taking into account the expression for kernel
$ T_{\Delta}^{1/2}(\vec{x},\vec{y}) $,
    one can write down the difference of traces of the square roots of 
$ T_{\MM} $
    and of the Laplace operator
\begin{align}
\nonumber
    E(\MM) = \Tr T_{\MM}^{1/2} - \Tr \Delta^{1/2}  
    &= \frac{1}{2\pi i}\int_{0}^{\infty} \bigl(
	\bb_{\alpha\beta}(\rho+i0)(D_{\rho-i0}^{\alpha},D_{\rho+i0}^{\beta}) -\\
\label{EM}
	&-\bb_{\alpha\beta}(\rho-i0)(D_{\rho+i0}^{\alpha},D_{\rho-i0}^{\beta})
    \bigr) \sqrt{\rho} \,d\rho.
\end{align}
    Transformation
(\ref{GD})
    allows one to rewrite the scalar products
$ (D_{\rho-i0}^{\alpha}, D_{\rho+i0}^{\beta}) $
    as derivatives of the elements of the matrix of boundary values
$ \Gamma_{\mu}^{\alpha\beta} $;
    indeed,
\begin{equation}
\label{DD}
    (D_{\bar{\mu}}^{\alpha}, D_{\lambda}^{\beta})
	= \frac{\Gamma_{\mu}^{\alpha\beta}
	    -\Gamma_{\lambda}^{\alpha\beta}}{\mu-\lambda} .
\end{equation}
    This yields
\begin{equation*}
    (D_{\bar{\mu}}^{\alpha}, D_{\mu}^{\beta}) = \frac{d}{d\mu}
	\Gamma_{\mu}^{\alpha\beta} ,
\end{equation*}
    and then, taking into account solution
(\ref{bexpr}),
    we arrive at the following expression for trace
(\ref{EM})
\begin{multline}
\label{EMM}
    E(\MM) = \frac{1}{2\pi i} \int_{0}^{\infty} \bigl(
    \frac{d\Gamma_{\mu}^{\alpha\beta}}{d\mu}
	(\MM-\Gamma_{\mu})^{-1}_{\alpha\beta} \Bigr|_{\mu=\rho+i0} -\\
    - \frac{d\Gamma_{\mu}^{\alpha\beta}}{d\mu}
	(\MM-\Gamma_{\mu})^{-1}_{\alpha\beta} \Bigr|_{\mu=\rho-i0}
	\bigr) \sqrt{\rho} \, d\rho .
\end{multline}
    When applied to norm
(\ref{NormOmOm})
    the functional calculus of operator
$ T_{\MM} $
    yields the following equation
\begin{multline}
\label{NormInt}
    \frac{\braket{\Omega_{\MM}}{\Omega_{\MM}}}{
	\braket{\Omega_{\Delta}}{\Omega_{\Delta}}}
     = \exp\bigl\{\frac{1}{8\pi i} \int_{0}^{\infty} \bigl(
    \frac{d\Gamma_{\mu}^{\alpha\beta}}{d\mu}
	(\MM-\Gamma_{\mu})^{-1}_{\alpha\beta} \bigr|_{\mu=\rho+i0} -\\
    - \frac{d\Gamma_{\mu}^{\alpha\beta}}{d\mu}
	(\MM-\Gamma_{\mu})^{-1}_{\alpha\beta} \bigr|_{\mu=\rho-i0}
	\bigr) \ln\rho \, d\rho \bigr\}.
\end{multline}
    Thus, the admissibility condition for functional
$ \Omega_{\MM} $
    represented by expression
(\ref{OmMT})
    results in the absence of discrete spectrum for
$ T_{M} $
    and in the convergence of integrals
(\ref{EMM}) and
(\ref{NormInt}).

\section{Convergence of the integrals}
    The main role in the calculation of integrals
(\ref{EMM}) and
(\ref{NormInt})
    is played by matrix
$ \Gamma_{\mu}^{\alpha\beta} $.
    Initially this object is defined as the matrix of the second coefficients
    of singularity expansions of the analytic deficiency vectors at
    points
$ \vec{x}_{1},\ldots \vec{x}_{N} $.
    Another definition can be extracted from relation
(\ref{DD}) ---
    the latter can be treated as an equation for matrix
$ \Gamma_{\mu}^{\alpha\beta} $
    with its solutions fixed up to a constant \emph{w.\,r.\,t.} the spectral
    parameter.
    This is the way in which the
$ Q $-functions of the theory of singular perturbations are introduced
\cite{AK}.
    In subsequent calculations we shall not worry about
    the genesis of
$ \Gamma_{\mu}^{\alpha\beta} $
    provided that its off-diagonal elements vanish at the infinity of
$ \mu $.

    In this work we restrict to the case of
$ N=2 $
    of the interaction of field
$ B $
    with two external particles. In this setup matrix
$ \Gamma_{\mu}^{\alpha\beta} $
    can be diagonalized by a single transformation for all
$ \mu $,
    and most of the calculations can be performed without the numerical
    analysis.
    For the case of a scalar field interacting with two point-like sources
    the matrix in question looks as follows
\begin{equation}
\label{Gscal}
    \Gamma_{\mu}^{nn'} = \begin{pmatrix}
    i\sqrt{\mu} & \frac{e^{i\sqrt{\mu}r}}{r} \\
    \frac{e^{i\sqrt{\mu}r}}{r} & i\sqrt{\mu}
    \end{pmatrix} , \quad 
\end{equation}
    where
$ r = |\vec{x}_{2}-\vec{x}_{1}| $ is the distance between points
$ \vec{x}_{1} $ and
$ \vec{x}_{2} $.
    The transverse vector field was studied in
\cite{TB2}.
    Borrowing the notation
$ \alpha = (n,m) $,
$ \beta = (n',m') $,
    and, agreeing that index
$ n $
    labels rows and
$ n' $
    labels columns,
$ \Gamma_{\mu}^{\alpha\beta} $
    can be written in the form of the following block matrix
\begin{equation}
\label{GTB}
\Gamma_{\mu}^{nm,n'm'}
    = \begin{pmatrix}
    i\sqrt{\mu} I_{mm'} & \frac{1}{r}\ww(\sqrt{\mu} r) (3\EE_{mm'}-I_{mm'}) \\
    \frac{1}{r}\ww(\sqrt{\mu} r) (3\EE_{mm'}-I_{mm'}) & i\sqrt{\mu} I_{mm'} 
\end{pmatrix} .
\end{equation}
    Here
\begin{equation*}
    I_{mm'} = \delta_{mm'},
    \quad \EE_{mm'} = \frac{(\vec{x}_{2}-\vec{x}_{1})_{m}
	(\vec{x}_{2}-\vec{x}_{1})_{m'}}{|\vec{x}_{2}-\vec{x}_{1}|^{2}},
\end{equation*}
    and function
$ w $
    comes from the coefficient in
(\ref{Dtr}).
    As long as elements of
(\ref{Gscal}) and
(\ref{GTB})
    are scalar products, the equations above can be easily generalized
    to an arbitrary value of
$ N $.
    It is sufficient to set the number of rows and columns to
$ N $,
    and to replace
$ \vec{x}_{2}-\vec{x}_{1} $
    with 
$ \vec{x}_{n}-\vec{x}_{n'} $
    and distances
$ r=|\vec{x}_{2}-\vec{x}_{1}| $ with
$ r_{nn'}=|\vec{x}_{n}-\vec{x}_{n'}| $
    in the off-diagonal block elements.

\subsection{Finiteness of the difference of the eigenvalues}
    Let us find the conditions on matrix
$ \MM $
    for integral
(\ref{EMM})
    to be finite. The divergence of this integral is mainly connected
    to the behavior of the integrand function at infinity.
    In general case we should take into account
    the action of projectors as in
(\ref{bexprp}).
    Namely, let matrices 
$ \Gamma_{\mu}^{mn} $ and
$ \Gamma_{\mu}^{'mn} $
    be of the form
\begin{equation}
\label{GG}
    \Gamma_{\mu}^{mn} = i\sqrt{\mu}\delta_{mn} + G_{mn}, \quad
    \Gamma_{\mu}^{'mn} = \frac{i}{2\sqrt{\mu}} (\delta_{mn} + G'_{mn}),
    \quad m,n = 1,\ldots M,
\end{equation}
    where 
$ G_{mn} $ and
$ G'_{mn} $
    have the following expansion at infinity
\begin{equation}
\label{GGexp}
    G_{mn} = \frac{C_{mn}^{kl} e^{i\sqrt{\mu}r_{kl}}}{\sqrt{\mu}}
	+ \OO_{mn}(\frac{1}{\mu}) ,\quad
    G'_{mn} = \frac{ir_{kl} C_{mn}^{kl} e^{i\sqrt{\mu}r_{kl}}}{\sqrt{\mu}}
	+ \OO'_{mn}(\frac{1}{\mu}) .
\end{equation}
    Assuming summation in the indices of
$ r_{mn} $
    in the RHSs of these expressions,
    one can see that
    the setup
(\ref{GG}),
(\ref{GGexp})
    covers both the bare matrix
(\ref{GTB})
    and the action of an arbitrary projector onto it.
    Let us write down the following estimates
\begin{equation}
\label{DetMG}
    \Det(\MM-\Gamma) = (-i\sqrt{\mu})^{M} + \Tr(\MM-\Gamma)(-i\sqrt{\mu})^{M-1}
	+ \OO(-i\sqrt{\mu})^{M-2} ,
\end{equation}
\begin{multline}
    (\MM-\Gamma)_{mn}^{-1} = \Det^{-1}(\MM-\Gamma)
    \bigl[ \bigl((-i\sqrt{\mu})^{M-1}
    +\Tr(\MM-G)(-i\sqrt{\mu})^{M-2}\bigr)\delta_{mn} -\\
\label{MG1}
    - (\MM-G)_{mn}(-i\sqrt{\mu})^{M-2} + \OO_{mn}(-i\sqrt{\mu})^{M-3}
\bigr] .
\end{multline}
    Upon multiplying
(\ref{GG}) by
(\ref{MG1})
    and summing up over the indices we find
\begin{multline}
    \Tr\bigl((\MM-\Gamma)^{-1}\frac{d\Gamma}{d\mu}\bigr)
    = \frac{i}{2\sqrt{\mu}}\Det^{-1}(\MM-\Gamma) \bigl[
(M+G'_{mm})(-i\sqrt{\mu})^{M-1} +\\
+ \bigl((M-1)\Tr(\MM-G) +\Tr(\MM-G)\Tr G' + \Tr(\MM-G)G'
\bigr)(-i\sqrt{\mu})^{M-2} +\\
\label{TrMG}
    + \OO(-i\sqrt{\mu})^{M-3} \bigr] .
\end{multline}
    The coefficient of
$ (-i\sqrt{\mu})^{M-2} $
    in this expression,
\begin{equation*}
    \omega(\mu) = (M-1)\Tr(\MM-G) +\Tr(\MM-G)\Tr G' + \Tr(\MM-G)G'
\end{equation*}
    has the following behavior at infinity
\begin{equation}
\label{omexp}
    \omega(\mu) = (M-1)\Tr(\MM-G)
+ \frac{C_{mn}e^{i\sqrt{\mu}r_{mn}}}{\sqrt{\mu}} + \OO(\frac{1}{\mu}) ,
\end{equation}
    where
$ C_{mn}e^{i\sqrt{\mu}r_{mn}} $
    is, again, linear combination of exponents with different periods.
    Substituting
(\ref{TrMG}) and
(\ref{DetMG}) into
(\ref{EMM})
    we come to
\begin{multline}
    E(\MM) =\\= \frac{1}{2\pi} \int_{0}^{\infty} \Bigl[
\frac{(M+G_{mm}^{'+})(-i{\sqrt{\rho}})^{M-1}+\omega^{+}(-i\sqrt{\rho})^{M-2}
    +\OO(-i\sqrt{\rho})^{N-3}}{(-i\sqrt{\rho})^{M}
	+\Tr(\MM-G^{+})(-i\sqrt{\rho})^{M-1} + \OO(-i\sqrt{\rho})^{M-2}} +\\
\label{EMI}
+\frac{(M+G_{mm}^{'-})(i{\sqrt{\rho}})^{M-1}+\omega^{-}(i\sqrt{\rho})^{M-2}
    +\OO(i\sqrt{\rho})^{N-3}}{(i\sqrt{\rho})^{M}
	+\Tr(\MM-G^{-})(i\sqrt{\rho})^{M-1} + \OO(i\sqrt{\rho})^{M-2}} 
\Bigr] d\rho .
\end{multline}
    Here
\begin{equation*}
    G_{mn}^{\pm} = G_{mn}(\pm \sqrt{\rho}) ,\quad
    G_{mn}^{'\pm} = G'_{mn}(\pm \sqrt{\rho}) ,\quad
    \omega^{\pm} = \omega(\pm \sqrt{\rho}),
\end{equation*}
    and the positive sign in front of the second term follows from the relation
\begin{equation*}
    \frac{\sqrt{\rho}}{\sqrt{\rho-i0}} =
    \frac{\sqrt{\rho+i0}}{\sqrt{\rho-i0}} =-1.
\end{equation*}
    Now one can bring the two terms to a common denominator and obtain
    the following integral
\begin{multline}
\label{Nom}
    E(\MM) = \frac{1}{2\pi} \int_{0}^{\infty} \Bigl[
	\frac{i\Tr(G^{'+} - G^{'-})\rho^{M-1/2}}{
    \rho^{M} + i\Tr(G^{-}-G^{+})\rho^{M-1/2} + \OO(\rho^{M-1}) } + \\
    + \frac{
	\bigl(2M\Tr\MM - \omega^{+}-\omega^{-} -M\Tr(G^{+}+G^{-})
	    -\Tr G^{'+} \Tr G^{-}
	\bigr)\rho^{M-1}
    }{\rho^{M} + i\Tr(G^{-}-G^{+})\rho^{M-1/2} + \OO(\rho^{M-1}) } + \\
    + \frac{ \bigl(\Tr\MM \Tr(G^{'+}+G^{'-}) -\Tr G^{'-} \Tr G^{+}
    \bigr)\rho^{M-1} + \OO(\rho^{M-3/2})
    }{\rho^{M} + i\Tr(G^{-}-G^{+})\rho^{M-1/2} + \OO(\rho^{M-1}) } \Bigr]
	\,d\rho .
\end{multline}
    Substituting estimates
(\ref{GGexp}),
(\ref{omexp})
    for
$ G_{mn} $,
$ G'_{mn} $ and
$ \omega $
    into this expression,
    one can see that the integral of the first term
    in the square brackets
    converges as the sum of integrals of the type
\begin{equation}
\label{Csin}
    \int_{0}^{\infty} \sin(\sqrt{\rho}r_{mn}) \frac{d\rho}{\rho} .
\end{equation}
    Meanwhile, the integral of the second term contains a logarithmic divergence
    proportional to
$ \Tr\MM $
\begin{multline}
\label{logdiv}
    \frac{1}{2\pi}\int_{0}^{\infty} \frac{(2M\Tr\MM - \omega^{+}-\omega^{-})
    \rho^{M-1}}{
    \rho^{M} + i\Tr(G^{-}-G^{+})\rho^{M-1/2} + \OO(\rho^{M-1})} \,d\rho =\\
=    \frac{1}{2\pi}\int_{0}^{\infty} \frac{2\Tr\MM\rho^{M-1}
	-\tilde{C}_{mn}\sin(\sqrt{\rho}r_{mn})\rho^{M-3/2} + \OO(\rho^{M-2})
    }{
    \rho^{M} + i\Tr(G^{-}-G^{+})\rho^{M-1/2} + \OO(\rho^{M-1})} \,d\rho .
\end{multline}
    From this it follows that the necessary and sufficient condition for the
    convergence of
(\ref{EMI})
    is the vanishing of the trace of
$ \MM $
\begin{equation}
\label{TrM}
    \Tr \MM = 0.
\end{equation}
    A similar calculation for the scalar field and matrix
(\ref{Gscal})
    shows that, besides condition
(\ref{TrM}),
    the finiteness of the corresponding eigenvalue difference also requires
    the convergence of the integral
\begin{equation*}
    \frac{i}{2\pi} \int_{0}^{\infty} \frac{\Tr(G^{'-}-G^{'+})}{
	\sqrt{\rho}} \,d\rho ,
\end{equation*}
    which in this case does not boil down to that of integral
(\ref{Csin}).

    The above calculation can also be used to verify the
    convergence of the difference
    of logarithm traces
(\ref{NormInt}).
    The calculation is carried out identically, with the denominator in
(\ref{Nom})
    acquiring an additional multiplier
$ \sqrt{\rho} $
    and the numerator
    multiplied by
$ \ln\rho $.
    These changes relax condition
(\ref{TrM}),
    that is, fraction
(\ref{NormInt})
    stays finite for an arbitrary matrix
$ \MM $
    not producing any discrete spectrum for the corresponding operator
$ T_{\MM} $.

\subsection{Non-negativity of the quadratic forms}
    As was shown above matrix
$ \MM $,
    defining an admissible self-adjoint extension of the operator
$ \Delta_{\{\vec{x}_{n}\}} $,
    should have a zero trace and should not produce discrete eigenvalues.
    Let us see how these requirements combine each with other.
    At first consider the limiting case
$ r \to \infty $
($ r_{nn'} \to \infty $ for 
$ N>2 $).
    Matrices
$ \Gamma_{\mu} $,
$ \Gamma'_{\mu} $
    then are proportional to the identity matrix
\begin{equation}
\label{GI}
    \Gamma_{\mu} = i\sqrt{\mu} I ,\quad
    \Gamma'_{\mu} = \frac{i}{2\sqrt{\mu}} I.
\end{equation}
    Since the value of
$ i\sqrt{\mu} $
    spans only the left half-plane of
$ \CC $, the poles of function
\begin{equation*}
    \bb(\mu) = (\MM - \Gamma_{\mu})^{-1}
	= (\MM - i\sqrt{\mu}I)^{-1}
\end{equation*}
    are generated by the negative eigenvalues of matrix
$ \MM $.
    It is obvious that the requirements of a zero trace of
$ \MM $
    and the absence of poles in
$ \bb(\mu) $
    can coexist only when matrix
$ \MM $
    equals to zero,
    and result in a zero value for
$ E(\MM) $.
    We would like to note that in this case the self-adjoint extension
    corresponding to
$ \MM=0 $
    is not the Laplace operator.

    Let us turn to the case of a finite distance
$ r $ ($ r_{nn'}$ for
$ N>2 $).
    It is clear that the absence of the discrete spectrum of
$ T_{\MM} $
    requires its quadratic form to be non-negative
\begin{equation*}
    0 \leq q_{\MM}(B) , \quad B\in \DD[T_{\MM}] .
\end{equation*}
    Now consider the following partial ordering on the set of self-adjoint
    extensions:
    we say that
$ T_{\MM_{1}} \leq T_{\MM_{2}} $,
    when
$ \DD[T_{\MM_{2}}] \subset \DD[T_{\MM_{1}}] $
    and
\begin{equation}
\label{ordering}
    q_{\MM_{1}}(B) \leq q_{\MM_{2}}(B) , \quad B\in\DD[T_{\MM_{2}}] .
\end{equation}
    The BKV theory states that the set of non-negative self-adjoint extensions
    of a closable symmetric operator with finite deficiency indices
    contains minimal
$ T_{\MM_{\text{K}}} $
    and maximal
$ T_{\MM_{\text{F}}} $
    extensions (conventionally named after M.~Krein and K.~Friedrichs,
    correspondingly). That is, for any non-negative extension
$ T_{\MM} $
    the following relations
\begin{equation*}
    T_{\MM_{\text{K}}} \leq T_{\MM}, \quad
    T_{\MM} \leq T_{\MM_{\text{F}}}
\end{equation*}
    take place. Besides this, ordering
(\ref{ordering})
    is monotonous \textit{w.\,r.\,t.}
$ \MM $
    in the sense of the ordering of finite-dimensional matrices.
    That is, if
$ T_{\MM_{1}} \leq T_{\MM_{2}} $,
    then
$ \DD(\MM_{2}) \subset \DD(\MM_{1}) $ and
\begin{equation*}
    (\xi, \MM_{1} \xi) \leq (\xi,\MM_{2}\xi), \quad \xi\in \DD(\MM_{2}) ,
\end{equation*}
    where we take into account that matrices
$ \MM_{1} $ and
$ \MM_{2} $
    might be defined on different domains (subspaces of
$ \CC^{3N} $).
    Taking into account the vanishing of traces of
$ \MM_{1} $,
$ \MM_{2} $
    and of their difference, this means that the convergence of integral
(\ref{EMM})
    for two matrices
$ \MM_{1} $ and
$ \MM_{2} $
    yields their coincidence on the intersection of their domains,
\begin{equation*}
    \MM_{1} \xi = \MM_{2}\xi ,\quad \xi\in\DD(\MM_{2})\cap \DD(\MM_{1}) .
\end{equation*}
    In our case, the maximal extension is the Laplace operator
$ \Delta $
    which corresponds to matrix
$ \MM_{\text{F}} $
    defined on the null set,
    and it produces the ground state
$ \Omega_{\Delta} $
    of the free theory.
    The minimal extension corresponds to matrix
\begin{equation*}
    \MM_{\text{K}} = \Gamma_{\mu} \bigr|_{\mu=0}
    = \begin{pmatrix}
    0 & -\frac{3}{4r} (3\EE_{mm'}-I_{mm'}) \\
    -\frac{3}{4r} (3\EE_{mm'}-I_{mm'}) & 0
\end{pmatrix} , \quad N=2,
\end{equation*}
    which is admissible \textit{w.\,r.\,t.} the convergence of integrals
(\ref{EMM}),
(\ref{NormInt}).
    In the scalar case, matrix
$ \MM_{K} $
    looks simpler
\begin{equation*}
    \MM_{\text{K}} = \begin{pmatrix}
    0 & \frac{1}{r} \\
    \frac{1}{r} & 0
    \end{pmatrix} , \quad N=2 
\end{equation*}
    and has a zero trace as well.
    Thus, we may conclude that all self-adjoint extensions corresponding
    to admissible eigenstates
$ \Omega_{\MM} $
    of the free Hamiltonian
    are generated by the matrix of Krein's extension restricted by
    orthogonal projectors
$ \PP_{\alpha}^{m} $
    onto subspaces which preserve the condition of the zero trace
\begin{equation*}
    \PP_{\alpha}^{m} \MM_{\text{K}}^{\alpha\beta} \PP_{\beta}^{m} = 0.
\end{equation*}
    We also would like to note that for the cases
$ r\to\infty $
    or
$ N=1 $
    described by matrix
(\ref{GI}),
    the Krein's extension corresponds to a zero matrix
$ \MM_{\text{K}} = 0 $.
    And its restriction to any subset, when matrix
$ \Gamma_{\mu} $
    is proportional to identity, produces a vanishing energy difference
(\ref{EMM}).
    That is, for the case
$ N=1 $
    all admissible ground states
$ \Omega_{\MM} $
    are not favoured
    to those of the free theory
    from the point of view of minimization of energy.

\section{Calculation of the eigenvalue difference}
\subsection{Krein's extension}
    Let us calculate integral
$ E(\MM_{\text{K}}) $
    for the matrix corresponding to the minimal extension.
    Consider an orthonormal basis
($ \vec{r} $, $ \vec{p} $, $ \vec{q} $)
    in the 3-dimensional space
\begin{equation*}
    \vec{r} = \frac{\vec{x}_{2}-\vec{x}_{1}}{|\vec{x}_{2}-\vec{x}_{1}|},\quad
    \vec{p} \cdot \vec{r} =0, \quad
    \vec{q} \cdot \vec{r} =0, \quad
    \vec{p} \cdot \vec{q} =0.
\end{equation*}
    These vectors are eigenvectors of the block elements of
(\ref{GTB}),
    indeed
\begin{equation*}
    (3J -I) \vec{r} = 2\vec{r},\quad
    (3J -I) \vec{p} = -\vec{p},\quad
    (3J -I) \vec{q} = -\vec{q}.
\end{equation*}
    It is not difficult to see that a change of basis
\begin{equation}
\label{vch}
    \ee_{mn} \to \tilde{\ee}_{\pm}^{\{r,p,q\}} = \frac{1}{\sqrt{2}}
	(\ee_{m1}\pm \ee_{m2})\{r_{m},p_{m},q_{m}\}
\end{equation}
    diagonalizes 
$ \Gamma_{\mu}^{nm,n'm'} $
    for any
$ \mu $,
    with the following eigenvalues
\begin{equation*}
    \gamma_{\pm}^{r}(\mu) = i\sqrt{\mu} \pm \frac{2}{r}\ww(\sqrt{\mu}r) , \quad
    \gamma_{\pm}^{\{p,q\}}(\mu) = i\sqrt{\mu} \mp \frac{1}{r}\ww(\sqrt{\mu}r).
\end{equation*}
    Function
$ \ww(t) $
    is defined in
(\ref{omt}).
    As a consequence, change
(\ref{vch})
    also diagonalizes matrix
$ \MM_{\text{K}} = \Gamma_{\mu}|_{\mu=0} $
    with eigenvalues
\begin{equation*}
    \gamma_{\text{K}\pm}^{r}(\mu) = \mp \frac{3}{2r}, \quad
    \gamma_{\text{K}\pm}^{\{p,q\}}(\mu) = \pm \frac{3}{4r} ,
\end{equation*}
    and matrix
$ \Gamma'_{\mu} $
    with eigenvalues
\begin{equation*}
    \gamma'{}_{\pm}^{r}(\mu) = \frac{i}{2\sqrt{\mu}}
	\bigl(1\mp 2\ww'(\sqrt{\mu}r)\bigr), \quad
    \gamma'{}_{\pm}^{\{p,q\}}(\mu) = \frac{i}{2\sqrt{\mu}}
	\bigl(1\pm \ww'(\sqrt{\mu}r)\bigr).
\end{equation*}
    In this way integral
(\ref{EMM})
    for matrix
$ \MM_{\text{K}} $
    decomposes into 6 terms, of which only 4 differ from each other.
    In terms of a dimensionless variable
$ t=\sqrt{\rho}r $
    the integral reads as follows
\begin{equation*}
    E(\MM_{\text{K}}) = \frac{1}{r}
	\sum_{\substack{\chi=1,1,-2,\\-1,-1,2}} \frac{1}{2\pi}
	\int_{0}^{\infty} \RE \frac{\chi \ww'(t)-1}{
	\frac{3}{4}\chi +it+\chi \ww(t)} t\,dt .
\end{equation*}
    A numerical calculation of the above sum gives
    an approximate value
$ -0.63 $,
    which shows that ground state
$ \Omega_{\MM_{\text{K}}} $
    is more favoured from the perspective of energy minimization, than
    ground state
$ \Omega_{\Delta} $
    of the free theory.
    A similar calculation for the case of interaction of a scalar field
    with two point-like sources (external particles) gives a
    dependence of the following type
\begin{equation*}
    E(\MM_{\text{K}}) = -0.33 \frac{1}{r} .
\end{equation*}

\subsection{The connected component}
    Let us show that the set of admissible matrices for
$ N=2 $
    possesses a connected component. We will seek it in the following
    rotation invariant way. Consider 3 vectors
\begin{align*}
    D_{\mu}^{r,\theta} &= r_{m} (\cos\theta D_{\mu}^{m1}
	    +\sin\theta D_{\mu}^{m2}) ,\\
    D_{\mu}^{p,\theta} &= p_{m} (\cos\theta D_{\mu}^{m1}
	    +\sin\theta D_{\mu}^{m2}) ,\\
    D_{\mu}^{q,\theta} &= q_{m} (\cos\theta D_{\mu}^{m1}
	    +\sin\theta D_{\mu}^{m2}) ,
\end{align*}
    where
$ \theta $
    is a dimensionless parameter.
    It is not difficult to see that
$ D_{\mu}^{s,\theta} $,
$ s=r,p,q $,
    constructed as linear transforms of
$ D_{\mu}^{mn} $,
    are themselves analytic deficiency vectors.
    This allows us to construct a set of 
$ \theta $-dependent
    resolvents of self-adjoint
    extensions of operator
$ \Delta_{\{\vec{x}_{n}\}} $.
    Scalar products of vectors
$ D_{\mu}^{s,\theta} $
    are readily calculated
\begin{align*}
    (\mu-\lambda) (D_{\bar{\mu}}^{r,\theta}, D_{\lambda}^{r,\theta}) &=
	i\sqrt{\mu} - i\sqrt{\lambda} +2\sin 2\theta \bigl(
    \frac{1}{r}\ww(\sqrt{\mu}r) - \frac{1}{r}\ww(\sqrt{\lambda}r)\bigr) ,\\
    (\mu-\lambda) (D_{\bar{\mu}}^{p,\theta}, D_{\lambda}^{p,\theta}) &=
	i\sqrt{\mu} - i\sqrt{\lambda} -\sin 2\theta \bigl(
    \frac{1}{r}\ww(\sqrt{\mu}r) - \frac{1}{r}\ww(\sqrt{\lambda}r)\bigr) ,\\
    (\mu-\lambda) (D_{\bar{\mu}}^{q,\theta}, D_{\lambda}^{q,\theta}) &=
	i\sqrt{\mu} - i\sqrt{\lambda} -\sin 2\theta \bigl(
    \frac{1}{r}\ww(\sqrt{\mu}r) - \frac{1}{r}\ww(\sqrt{\lambda}r)\bigr) ,\\
    (D_{\bar{\mu}}^{r,\theta}, D_{\lambda}^{p,\theta}) &=
    (D_{\bar{\mu}}^{r,\theta}, D_{\lambda}^{q,\theta}) =
    (D_{\bar{\mu}}^{p,\theta}, D_{\lambda}^{q,\theta}) = 0 
\end{align*}
    and lead to the following diagonal solution of
(\ref{DD})
\begin{equation}
\label{Gtheta}
    \Gamma_{\mu}^{\theta} = i\sqrt{\mu} I +\sin 2\theta \,
	\frac{\ww(\sqrt{\mu}r)}{r}
\begin{pmatrix} 2 & 0&0 \\ 0& -1 &0 \\ 0& 0& -1
\end{pmatrix}.
\end{equation}
    This matrix satisfies conditions
(\ref{GG}), (\ref{GGexp})
    of the convergence of integral
(\ref{EMM}),
    and its construction guarantees that the determinant of matrix
\begin{equation*}
    \MM(\theta) -\Gamma_{\mu}^{\theta}
	= \Gamma_{0}^{\theta} - \Gamma_{\mu}^{\theta}
    = -i\sqrt{\mu}I - \sin 2\theta \bigl(\frac{3}{4r}+\frac{w(\sqrt{\mu}r)}{r}
    \bigr)
\begin{pmatrix} 2 & 0&0 \\ 0& -1 &0 \\ 0& 0& -1
\end{pmatrix}
\end{equation*}
    does not turn zero. This shows that the resolvent
\begin{equation*}
    R_{\mu}^{\theta} = R_{\mu} + D_{\mu}^{s,\theta}
	(\MM(\theta)-\Gamma_{\mu}^{\theta})_{ss'}^{-1}
	(D_{\bar{\mu}}^{s',\theta}, \,\cdot\,),
    \quad s=r,p,q
\end{equation*}
    corresponds to a self-adjoint extension with a finite value of integral
$ E(\MM(\theta)) $.
    Substituting
(\ref{Gtheta})
    into
(\ref{EMM})
    we obtain the following expression
\begin{multline*}
    E(\MM(\theta)) = \frac{1}{2\pi r}
	\int_{0}^{\infty} \RE \Bigl(
    \frac{2\sin2\theta \,\ww'(t)-1}{\frac{3}{2}\sin2\theta +it
	+ 2\sin2\theta \,\ww(t)} +\\
    + 2\frac{\sin2\theta \, \ww'(t)+1}{\frac{3}{4}\sin2\theta -it
	+ \sin2\theta \,\ww(t)}
\Bigr) t\,dt .
\end{multline*}
\begin{center}
\vspace*{-0.5cm}
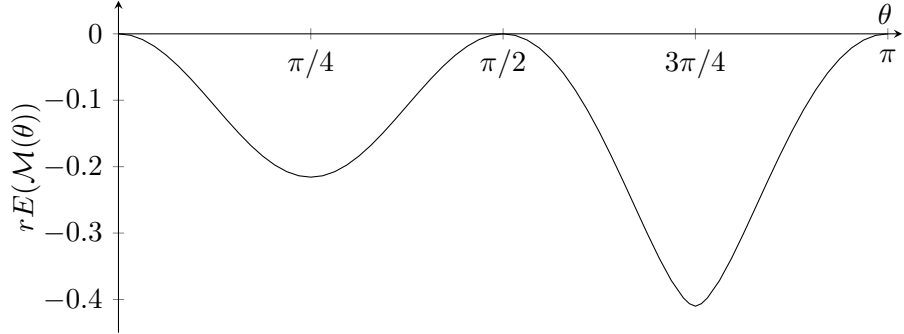
\begin{figure}
\begin{tikzpicture}
\begin{axis}[
xlabel={$\theta$},
ylabel={$ rE(\MM(\theta)) $},
xtick={0,0.78528,1.57056,2.35584,3.14159},
xticklabels={0,$ \pi/4$,$ \pi/2$, $ 3\pi/4$, $ \pi $},
ytick={-0.5,-0.4,-0.3,-0.2,-0.1,0,0.1},
axis x line=center,
axis y line=left,
xmin = 0, xmax=3.2, ymin=-0.45, ymax=0.05,
width=12cm,
height=6cm
]
\addplot[ mark=none,
color=black]
coordinates {(0,0)
(0.0499,-0.0022)
(0.09816,-0.0086)
(0.14724,-0.0189)
(0.19632,-0.0325)
(0.24540,-0.0489)
(0.29448,-0.0675)
(0.34356,-0.0876)
(0.39264,-0.1084)
(0.44172,-0.1293)
(0.49080,-0.1493)
(0.53988,-0.1679)
(0.58896,-0.1841)
(0.63804,-0.1975)
(0.68712,-0.2075)
(0.73620,-0.2137)
(0.78528,-0.2158)
(0.83436,-0.2137)
(0.88344,-0.2075)
(0.93252,-0.1975)
(0.98160,-0.1841)
(1.03068,-0.1679)
(1.07976,-0.1493)
(1.12884,-0.1293)
(1.17792,-0.1084)
(1.22700,-0.0876)
(1.27608,-0.0675)
(1.32516,-0.0489)
(1.37424,-0.0325)
(1.42332,-0.0189)
(1.47240,-0.0086)
(1.52148,-0.0022)
(1.57056,0)
(1.61964,-0.0023)
(1.66872,-0.0093)
(1.71780,-0.0211)
(1.76688,-0.0377)
(1.81596,-0.0589)
(1.86504,-0.0846)
(1.91412,-0.1143)
(1.96320,-0.1476)
(2.01228,-0.1839)
(2.06136,-0.2222)
(2.11044,-0.2616)
(2.15952,-0.3008)
(2.20860,-0.3382)
(2.25768,-0.3716)
(2.30676,-0.3976)
(2.33130,-0.4063)
(2.35584,-0.4101)
(2.38038,-0.4063)
(2.40492,-0.3976)
(2.45400,-0.3716)
(2.50308,-0.3382)
(2.55216,-0.3008)
(2.60124,-0.2616)
(2.65032,-0.2222)
(2.69940,-0.1839)
(2.74848,-0.1476)
(2.79756,-0.1143)
(2.84664,-0.0846)
(2.89572,-0.0589)
(2.94480,-0.0377)
(2.99388,-0.0211)
(3.04296,-0.0093)
(3.09204,-0.0023)
(3.14159,0)};
\end{axis}
\end{tikzpicture}
\caption{Dependence of $rE(\MM(\theta))$ on $ \theta $}
\label{fig:rE}
\end{figure}
\end{center}

    The dependence of the dimensionless quantity
$ rE(\MM(\theta)) $
    on
$ \theta $,
$ 0\leq\theta\leq\pi $
    is shown in Fig.~\ref{fig:rE}.
    Values
$ \theta = 0,\pi $ and
$ \theta=\frac{\pi}{2} $
    correspond to the Krein extension for
$ N=1 $,
    where the singular boundary condition resides either at point
$ \vec{x}_{1} $
$ (\theta=0,\pi) $,
    or at point
$ \vec{x}_{2} $
$ (\theta=\frac{\pi}{2}) $.
    As has been expected, energy difference
(\ref{EMM})
    for these solutions is equal to zero
$ E(\MM(\{0,\frac{\pi}{2},\pi\}))=0 $.
    Local minima
$ \theta=\frac{\pi}{4} $ and
$ \theta=\frac{3\pi}{4} $
    corresponds to boundary conditions for which the vector values of field
$ B $
    at points
$ \vec{x}_{1} $ and
$ \vec{x}_{2} $
    coincide
$ (\theta=\frac{\pi}{4}) $,
    or coincide up to a sign
$ (\theta=\frac{3\pi}{4}) $.
    Other values,
$ \theta \neq 0,\frac{\pi}{2},\pi $
    correspond to boundary conditions for which the values of the field at
    points
$ \vec{x}_{1} $ and
$ \vec{x}_{2} $
    are related by a coefficient of
$ \ctg\theta $
\begin{equation*}
    \lim_{\vec{x}\to\vec{x}_{1}} B_{\vec{x}} = \ctg\theta 
    \lim_{\vec{x}\to\vec{x}_{2}} B_{\vec{x}} ,
\end{equation*}
    that is, we arrive at a theory with action at a distance.

    We can see that the set of admissible matrices for the Hamiltonian
    of a transverse field possesses a connected component parametrized by
    a dimensionless quantity
$ \theta $.
    The energy difference
$ E(\MM(\theta)) $
    continuously varies from zero to a negative value,
    with change of
$ \theta $.
    This allows one to employ this model in the description of systems
    with dimensional transmutation.
    Indeed, the 2-particle eigenstates
    of the interacting fields, after renormalization,
    in the combined Shr\"odinger
    (for the field 
$ B $)
    and Fock (for the field
$ \psi $)
    representation
    can be constructed as the following sum
\begin{equation}
\label{OPB}
    \Omega(\psi, B) = \sum_{\sigma_{1},\sigma_{2}} \int_{\RR^{3}}
	\psi(\vec{x}_{1},\sigma_{1};\vec{x}_{2},\sigma_{2})
    \Psi_{\vec{x}_{1}}^{\sigma_{1}} \Psi_{\vec{x}_{2}}^{\sigma_{2}}
    \,\Omega_{\MM(\theta(\psi))}(B) \, d^{3}x_{1}\,d^{3}x_{2} .
\end{equation}
    Here
$ \psi $
    is a 2-particle wave function, and
$ \Psi_{\vec{x}}^{\sigma} $
    is the creation operator of the second field
    with quantum numbers
$ \sigma $
    at point
$ \vec{x} $.
    Parameter
$ \theta $
    can directly depend on the arguments of wave function
$ \psi $,
\textit{e.\,g.} on distance
$ |\vec{x}_{1}-\vec{x}_{2}| $.
    Any nontrivial dependence
\begin{equation*}
    \theta(|\vec{x}_{1}-\vec{x}_{2}|) \neq \text{const}
\end{equation*}
    implies the presence of a dimensional parameter and thus leads
    to dimensional transmutation in the theory.
    The specific form of that dependence is determined by the initial
    interaction and by the process of renormalization.
    The latter process includes the ``cloud'' of virtual particles
    and other non-perturbative effects. Its output is a Hamiltonian
    with many-particle states like
(\ref{OPB}),
    and the dependence of the generalized parameter
$ \theta $
    on the coordinates and quantum numbers of particles of field
$ \psi $.
 
\section*{Conclusion}
    We have considered the simplest solutions of the eigenstate equation
    of free quantum Hamiltonians of scalar and vector Coulomb-gauged
    fields. These solutions do not represent states of the free field,
    and, similarly to the objects of the finite-dimensional theory of
    singular perturbations, can be viewed as eigenstates of
    perturbed Hamiltonians of the free field. Another insight
    from the above theory is
    that the solutions in question are eigenstates of a certain asymptotically
    free system which has undergone the process of renormalization.

    The author is grateful to S.~Naboko for inspiring consultations on
    the Birman-Krein-Vishik theory and to P.~Bolokhov for valuable comments.

\end{document}